\documentclass[prl,twocolumn,amssymb,superscriptaddress]{revtex4}
\usepackage{graphicx}

\begin{document}

\title{Pairing and Vortex Lattices for Interacting Fermions in Optical Lattices with Large Magnetic Field}
\author{Hui Zhai}
\email{hzhai@mail.tsinghua.edu.cn}
\affiliation{Institute for Advanced Study, Tsinghua
University, Beijing, 100084, China}
\author{R. O. Umucal\i lar}
\affiliation{Department of Physics, Bilkent University, 06800,
Ankara, Turkey}
\author{ M. \"O. Oktel}
\affiliation{Department of Physics, Bilkent University, 06800,
Ankara, Turkey}
\date{\today}

\begin{abstract}
We study the structure of pairing order parameter for spin-$1/2$
fermions with attractive interactions in a square lattice under a
uniform magnetic field. Because the magnetic translation symmetry
gives a unique degeneracy in the single-particle spectrum, the
pair wave function has both zero and finite momentum components
co-existing, and their relative phases are determined by a
self-consistent mean-field theory. We present a microscopic
calculation that can determine the vortex lattice structure in the
superfluid phase for different flux densities. Phase transition
from a Hofstadter insulator to a superfluid phase is also
discussed.
\end{abstract}
\maketitle

Optical lattices and synthetic magnetic fields are two of major tools
to create strongly interacting many-body systems in cold atoms
\cite{review,fetter}. In conventional solid state materials,
accessible magnetic flux per unit cell $n_{\text{B}}$ is very
small, $n_{\text{B}}\ll 1$, even for the strongest magnetic field
attainable in laboratory ($\lesssim 45T$). Hence, as in most
conventional metals, the electron density $n$ is several orders larger than $n_{\text{B}}$ that the magnetic field can be treated
semi-classically; or as in the two-dimensional electron gases, $n\sim n_{\text{B}}\ll 1$, the density is so low that only the bottom of an
electron band is populated, and the effective mass approximation
is sufficient to account for the lattice effect. In cold atom
systems, because the magnetic field is synthetically generated by
rotation \cite{Cornell} or by engineering atom-light interactions
\cite{gauge_theory,NIST}, and the lattice spacing is of the order
of half a micron, one can access the regime $n\sim
n_{\text{B}}\sim 1$, where both the lattice and the magnetic field
should be treated on an equal footing and in a quantum-mechanical manner. Consequently, such a system
exhibits the famous Hofstadter butterfly single-particle spectrum
\cite{Hofstadter}.

For neutral atoms in lattices, the interaction is dominated by
on-site interactions as in the Hubbard model. Hereafter, we shall
refer to the model describing interacting cold atoms in optical
lattices with large magnetic field as the Hofstadter-Hubbard (HH)
model. Recently, many works have focused on the bosonic HH model
\cite{BHH}, which reveal a number of interesting phenomena,
including vortex lattice states and possible fractional quantum
Hall states. However, so far little attention has been paid to the
fermionic HH model.

The subject of this letter is the
properties of the paired superfluid phase in the fermionic HH
model with attractive interactions. For $n_{\text{B}}\sim 1$, the pairing problem differs from type-II superconductors in a fundamental way.
In type-II superconductors the separation between the vortices is
much larger than the size of Cooper pairs, hence one can locally
apply the BCS scenario to define a local order parameter
$\Delta({\bf r})$, and understand the vortex lattice by coupling
this ``coarse grained" order parameter to the magnetic field.
In the HH model considered here, magnetic field
modifies the single-particle dispersion in an important way. Despite a strong magnetic field,
there is always a well-defined Fermi surface and Bloch states in
the magnetic Brillouin zone (MBZ) in the Hofstadter model. Therefore, with attractive interaction BCS pairing
always occurs as an instability of the Fermi liquid. This enables us to reach the regime where the pair size is comparable to the distance between vortices, 
hence, any discussion of pairing
must include the effect of the magnetic field at the microscopic
level. We shall show that such a microscopic theory requires the
definition of an order parameter with multiple components, and will discuss how this order parameter naturally describes the configuration of vortices. The main points of our analysis are highlighted as follows.

({\bf 1}): We first review that for $n_{\text{B}}=p/q$, where $p$
and $q$ are co-prime integers, each single particle state in the
Hofstadter spectrum is $q$-fold degenerate due to magnetic
translation symmetry \cite{Zak,Niu}. This degeneracy enforces that
a comprehensive formulation of BCS theory in this case must
contain Cooper pairs with both zero and a set of finite momenta,
and treat them on an equal footing.

({\bf 2}): We show that the magnetic translation symmetry also
imposes relations between pairing order parameters of different
momentum. These relations are verified numerically by
self-consistently solving the BCS mean-field Hamiltonian.

({\bf 3}): The relative phases between different pairing order
parameters determined from self-consistent solutions can also be
understood from a more intuitive and simpler Ginzburg-Landau
argument.

({\bf 4}): We determine the structure of vortices in the
superfluid ground state using the information from ({\bf 2}). The
unit cell of the superfluid phase is enlarged to $q\times q$,
whose symmetry is lower than that of the original Hamiltonian.
Hence, the superfluid ground state has discrete degeneracy,
related to the symmetry of the vortex lattice.

({\bf 5}): For certain fermion densities, a critical interaction
strength is predicted for a quantum phase transition from a
Hofstadter insulator to a superfluid phase.

\begin{figure}[tbp]
\includegraphics[height=2.5in, width=3.0in]
{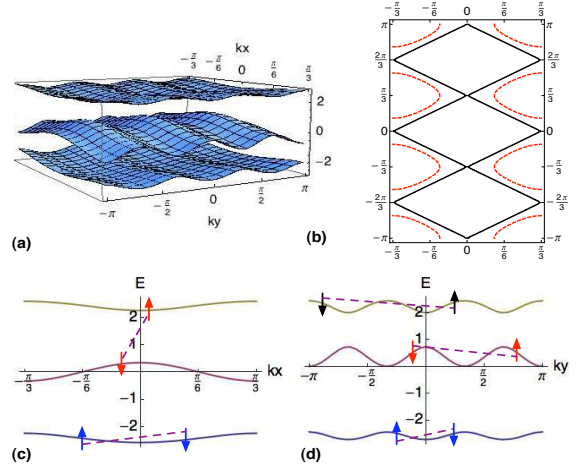} \caption{(Color online) (a) Three magnetic
bands for $p/q=1/3$. (b) The Fermi surface of a half-filled (black
solid line) and slightly away from half-filled (red dashed line)
system. (c) and (d): the band dispersion along $k_x$-direction
with $k_y=0$ (c) and along $k_y$ direction with $k_x=0$ (d).
Various possible pairings included in our BCS theory are also
illustrated in (c) and (d), which include intra- and inter-band
pairing (c), and pairing with non-zero center-of-mass momentum
(d). \label{model}}
\end{figure}

{\it The Model:} We consider a two-component Fermi gas in a
two-dimensional optical lattice potential so that an $s-$band
tight binding model accurately describes the dynamics. Both
components are coupled to the same gauge field $\vec{A}=(0,px/q)$
in the Landau gauge. Note that there is no Zeeman shift
associated with a synthetic magnetic field. The single-particle Hamiltonian is given by
\begin{equation}
H_{0}=-t\sum_{\langle ii'\rangle \sigma}\left(e^{i2\pi
A_{ii'}}c^\dag_{i\sigma}c_{i'\sigma}+\text{h.c.}\right),
\end{equation}
where $i=(i_x,i_y)$ labels the lattice sites, and $\langle
ii'\rangle$ represents all the nearest neighboring bonds. In the
Landau gauge, $A_{ii'}=pi_x/q$ if $i-i'$ is along $y$-direction
and $A_{ii'}=0$ if $i-i'$ is along $x$-direction. Let $T_{\hat{x}}$ and $T_{\hat{y}}$ be magnetic translations of one lattice spacing along $x-$ and $y-$ direction (see Ref. \cite{Zak,Niu} for definition). The Hamiltonian $H_0$ commutes with both $T_{\hat{x}}$ and $T_{\hat{y}}$,
but these two operators do not commute $T_{\hat{x}}T_{\hat{y}} = \exp\{i 2\pi
p/q\}T_{\hat{y}}T_{\hat{x}}$. One can choose the set of
commuting operators as $H_0$,$T_{\hat{y}}$, and $(T_{\hat{x}})^q
\equiv T_{q\hat{x}}$, in effect, enlarging the unit cell in real
space to contain $q$ sites in $x$-direction. Thus, MBZ becomes
$k_x\subset[-\pi/q,\pi/ q)$ and
$k_y\subset[-\pi, \pi)$. Denoting the common
eigenstate by $\psi_{n,k_x,k_y}$ and the
eigen-energy by $\epsilon_{nk_xk_y}$, the magnetic Bloch theorem
yields $T_{q\hat{x}}\psi_{n,k_x,k_y}=\exp\{iq
k_x\}\psi_{n,k_x,k_y}$, $T_{\hat{y}}\psi_{n,k_x,k_y}=\exp\{i
k_y\}\psi_{n,k_x,k_y}$, where $n$ is the band index. Thus, for
$T_{l\hat{x}}$, $l=1,\dots,q-1$,
$T_{l\hat{x}}\psi_{n,k_x,k_y}$ is a degenerate eigenstate of
$\psi_{n,k_x,k_y}$, and since
$T_{\hat{y}}(T_{l\hat{x}}\psi_{n,k_x,k_y})=\exp\{i (k_y+2\pi l
p/q)\}(T_{l\hat{x}}\psi_{n,k_x,k_y})$, we have the following
properties \cite{Zak,Niu}
\begin{eqnarray}
\psi_{n,k_x,k_y}(x+l \lambda,y)&\propto& \psi_{n, k_x,k_y+2\pi lp/q}(x,y),\nonumber\\
\epsilon_{n,k_x,k_y}&=&\epsilon_{n,k_x,k_y+2\pi lp/q}.
\end{eqnarray}
For $p/q=1/3$, the spectrum and Fermi surface shown in Fig.
\ref{model}(a,b) clearly display a three-fold degeneracy.

In addition to $H_0$, we consider the on-site interaction between
different spin components
\begin{equation}
H_{\text{int}}=U\sum_{i}c^\dag_{i\uparrow}c^\dag_{i\downarrow}c_{i\downarrow}c_{i\uparrow}.
\end{equation}
The Hamiltonian for the HH model discussed below is then given by
$H_{\text{HH}}=H_0+H_{\text{int}}$.

\begin{figure}[btp]
\includegraphics[height=1.7in,width=3.0in]
{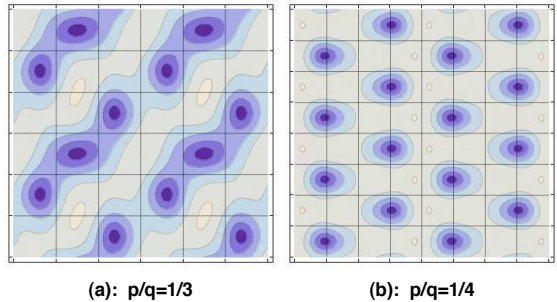} \caption{(Color online) Structure of the vortex
lattice found from self-consistently solving the BCS mean-field
Hamiltonian, where $n_{\text{B}}=1/3$ for (a) and
$n_{\text{B}}=1/4$ for (b). This is a contour plot of pairing
order parameter $\Delta({\bf r})$. The grey area means low
superfluid density and locates the center of vortex cores. The
intersection of vertical and horizontal straight lines indicates
lattice sites. In both plots we use $U=-5.5t$ and
$n_{\uparrow}=n_{\downarrow}=1/3$ for (a) and
$n_{\uparrow}=n_{\downarrow}=1/2$ for (b). \label{VL}}
\end{figure}

{\it Generalized BCS Theory:} We start
by diagonalizing the non-interacting Hamiltonian $H_0$. First we
relabel the sites to reflect the enlargement of the unit cell. For
a site $i=(i_x,i_y)$, we let $i_x=j_x q+\beta$, where $j_x$ is an
integer labelling the magnetic unit cell and $\beta=0,\dots,q-1$
denotes the $q$- inequivalent sites within each cell. So, magnetic unit
cells are uniquely labelled by $j_x$ and $j_y=i_y$. With this
notation we identify $c_{j\beta\sigma}=c_{i\sigma}$. Fourier
transformation in the variable $j$ yields $c_{{\bf
k}\beta\sigma}$, where ${\bf k}$ is limited inside the MBZ.
We now define a new set of operators $d_{{\bf k} n \sigma}$ through $c_{{\bf k}\beta
\sigma}=\sum_{n}g^n_\beta({\bf k})d_{{\bf k} n \sigma}$,  under
which $H_0$ becomes diagonalized as
\begin{equation}
H_0=\sum_{n{\bf k}\sigma}\epsilon_{n{\bf k}}d^\dag_{{\bf
k}n\sigma}d_{{\bf k}n\sigma},
\end{equation}
Note that diagonalization of $H_0$ is equivalent to solving Harper's
equation \cite{Hofstadter}, and
$g^n_\beta({\bf k})$ is the $\beta$th component of the $n$th
eigenvector of Harper's equation at wavevector ${\bf k}$  \cite{Hofstadter}.
$d^\dag_{{\bf k}n}$ is the operator that
creates a particle in the $n$th magnetic sub-band at wavevector
${\bf k}$. As an example, $\epsilon_{n{\bf k}}$ is plotted in Fig.
\ref{model}(a) for $p/q=1/3$. In terms of $d_{{\bf k}n\sigma}$,
$H_{\text{int}}$ becomes
\begin{eqnarray}
&&H_{\text{int}}=U\sum\limits_{\beta}\sum\limits_{{\bf
Q}}\sum\limits_{{\bf k},{\bf
k^\prime}}\sum\limits_{n_1,\dots,n_4}\nonumber\\ && g^{n_1
*}_{\beta} ({\bf \frac{Q}{2}+k})g^{n_2 *}_\beta({\bf
\frac{Q}{2}-k}) g^{n_3}_\beta({\bf
\frac{Q}{2}-k^\prime})g^{n_4}_{\beta}({\bf
\frac{Q}{2}+k^\prime})\times\nonumber\\ && d^\dag_{{\bf
\frac{Q}{2}+k},n_1,\uparrow} d^\dag_{{\bf
\frac{Q}{2}-k},n_2,\downarrow}d_{{\bf
\frac{Q}{2}-k^\prime},n_3,\downarrow}d_{{\bf
\frac{Q}{2}+k^\prime},n_4,\uparrow},\label{int}
\end{eqnarray}
where the momentum sum is restricted to the MBZ. We should focus
on the ``on-shell" Cooper processes with ${\bf k^\prime}=-{\bf
k}$. Importantly, due to the $q$-fold degeneracy, we not only
consider ${\bf Q}=0$ terms in Eq. (\ref{int}), but also need to
consider all the terms with ${\bf Q}=(0, 2\pi l p/q)$, where
$l=0,\dots,q-1$, since ${\bf -k}$ and ${\bf k+Q}$ also have the
same kinetic energy. Consequently, non-zero center-of-mass
momentum pairing needs to be included as well. Besides, intra-band
Cooper pairs have a non-vanishing coupling to the inter-band
Cooper pairs. For instance, in Eq. (\ref{int}), if $n_1=n_2$ but
$n_3\neq n_4$, the interaction coefficient is non-zero. Hence,
intra-band pairing must induce inter-band pairing. All these
pairing scenario under consideration are schematically illustrated
in Fig. \ref{model}(c-d), and a comprehensive BCS theory in this
problem must treat all these possibilities on an equal footing.
Therefore, we introduce totally $q^2$ order parameters
$\vec{\Delta}^{l}=(\Delta^l_0,\dots,\Delta^l_{q-1})$ given by
\begin{eqnarray}
\Delta^{l}_\beta = -U&\sum\limits_{n,n^\prime,{\bf
k}}&g^n_\beta({\bf k+\frac{Q}{2}})g^{n^\prime}_\beta({\bf
-k+\frac{Q}{2}}) \nonumber \\ &\times&\langle
d_{{\bf-k+\frac{Q}{2}},n^\prime,\downarrow}d_{{\bf
k+\frac{Q}{2}},n,\uparrow}\rangle,
\end{eqnarray}
where $l,\beta=0,\dots,q-1$. The site index $\beta$ denotes $q$
inequivalent sites along $x$-direction of each magnetic unit cell,
and the index $l$ represents the center-of-mass momentum of the
pair ${\bf Q}=(0, 2\pi l p/q)$, which represents the order
parameter modulation along $y$-direction. For instance, for
$p/q=1/3$, there are three different center-of-mass momenta, which
are ${\bf Q}_{l=0}=(0,0)$, ${\bf Q}_{l=1}=(0, 2\pi/3)$ and ${\bf
Q}_{l=2}=(0,4\pi/3)$. With $\Delta^l_\beta$, the mean-field Hamiltonian becomes
\begin{eqnarray}
&&H_{\text{MF}}=\sum\limits_{n{\bf k}\sigma}\epsilon_{n{\bf k}\sigma}d^\dag_{n{\bf k}\sigma}d_{n{\bf k}\sigma}-\sum\limits_{l,\beta}
\left\{\sum\limits_{n,n^\prime,{\bf k}} \right. \nonumber\\ && \left. \left(
\Delta^l_{\beta}g^{n *}_{\beta}({\bf k+\frac{Q}{2}})g^{n^\prime *}_{\beta}({\bf -k+\frac{Q}{2}})d^\dag_{{\bf k+\frac{Q}{2}},n,\uparrow}d^\dag_{{\bf -k+\frac{Q}{2}},n^\prime,\downarrow} \right.\right.\nonumber\\
&&\left.\left.+\text{h.c.}\right)+\frac{|\Delta^l_\beta|^2}{U}\right\}.
\label{MF}
\end{eqnarray}
The real space order parameter for site
$i=(i_x,i_y)$ is given by 
\begin{equation}
\Delta_i=\sum_{l=0}^{q-1}\Delta^l_{i_x
(\text{mod} \ \ q)}e^{i2\pi lp i_y/q},
\end{equation}
therefore the unit
cell in the superfluid phase is enlarged to $q\times q$ in real
space (see Fig. \ref{VL}).

{\it Solution to BCS Theory:} We start with $q^2$ random complex
numbers as initial $\Delta^{l}_{\beta}$ and iteratively solve the
BCS mean-field Hamiltonian [Eq. (\ref{MF})] until a
self-consistent solution is reached. We find for a convergent
solution, the $q^2$ order parameters are not completely
independent. In fact, these $q^2$ order parameters break up into
$q$ sets of $q$ order parameters with the same magnitude. Taking
$p/q=1/3$ or $1/4$ as examples, their relations are summarized in
Table \ref{relations}.

\begin{table}[htdp]
\begin{minipage}{1.5in}
\raggedleft
\begin{tabular}{|c|c|c|c|c|}
\hline
  $\Delta^l_{\beta}$  &$\beta=0$& $\beta=1$ &$\beta=2$\\
   \hline
  $l=0$ & $a$ & $b$ & $c$  \\
  \hline
  $l=1$ & $b e^{i\theta_1}$ & $c e^{i\theta_1}$ & $a e^{i\theta_1}$  \\
  \hline
    $l=2$ & $c e^{i\theta_2}$ & $a e^{i\theta_2}$ & $b e^{i\theta_2}$  \\
\hline
\end{tabular}
\end{minipage}
\raggedright
\begin{minipage}{1.8in}
\begin{tabular}{|c|c|c|c|c|}
\hline
  $\Delta^l_{\beta}$ & $\beta=0$ & $\beta=1$& $\beta=2$& $\beta=3$ \\
   \hline
  $l=0$ & a & $b$ & $a$ & $b$ \\
  \hline
  $l=1$  & $c$ & $d$ & $c$ & $d$  \\
  \hline
    $l=2$  & $b e^{i\theta_1}$ & $a e^{i\theta_1}$ & $b e^{i\theta_1}$ & $a e^{i\theta_1}$  \\
\hline
$l=3$ & $d e^{i\theta_2}$ & $c e^{i\theta_2}$ & $d e^{i\theta_2}$ & $c e^{i\theta_2}$  \\
\hline
\end{tabular}
\end{minipage}%
\caption{Pairing order parameters for $p/q=1/3$ (left) and
$p/q=1/4$(right). $a$, $b$, $c$, and $d$ denotes some complex
numbers depending on details, like the fermion density and
$U/t$.\label{relations}}
\end{table}

\begin{figure}[tbp]
\includegraphics[scale=.45]
{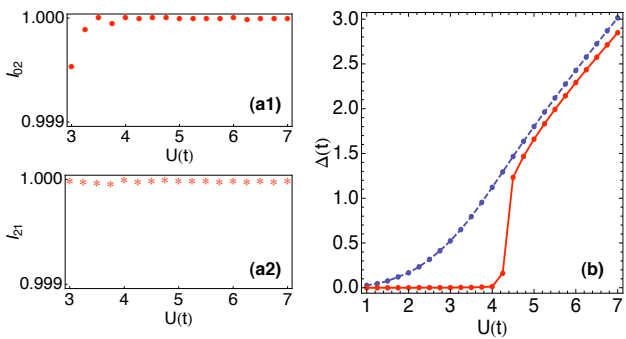} \caption{(Color online) (a) For $p/q=1/3$,
$I_{02}$ and $I_{21}$ (see text for definition) equal to unity
within numerical accuracy, which verifies the symmetry relations.
(b) Average superfluid order parameter $\Delta=\bar{\Delta}_i$ as
a function of $U$ for $p/q=1/3$, $n=1/3$ (red solid line) shows a
phase transition and $n=1/2$ (blue dashed line) does not.
\label{Phase}}
\end{figure}

These structures can be understood from the symmetry properties
discussed above. The system is invariant under translation by one
lattice site along $x$-direction and a simultaneous translation of
$k_y$ by $2\pi p/q$. Under this operation,
$\Delta_\beta^l\rightarrow\Delta_{\beta^\prime}^{l^\prime}$, where
$\beta^\prime=\beta+1\  \ (\text{mod} \ \ q)$ and $l^\prime=l+2\ \
(\text{mod} \  \ q)$, thus these two order parameters must be
equal up to a relative phase. To verify these relations, we show
in Fig. \ref{Phase}(a) that our numerical solutions satisfy
$I_{ll^\prime}=|\vec{\Delta}^{l^\prime}\Gamma\Delta^{l\dag}|/(|\vec{\Delta}^l||\vec{\Delta}^{l^\prime}|)=1$
for $l^\prime=l+2(\text{mod} \  \ q)$, where $\Gamma$ is a
$q\times q$ matrix with $\Gamma_{ij}=\delta_{i+1 (\text{mod} \  \
q),j}$. This symmetry imposed relation works for any $p/q$,
which implies that if $\Delta^0$ is non-zero, all
$\Delta^{2n(\text{mod} \  \ q)}$ are non-zero, i.e., zero and
finite-momentum component must co-exist.

The self-consistent solution also determines the relative phases.
For $p/q=1/3$, we find six degenerate solutions with
$(\theta_1,\theta_2)=(\pm 2\pi/3,\pm 2\pi/3),(0,\pm 2\pi/3),(\pm
2\pi/3,0)$; for $p/q=1/4$, we find $\theta_{1,2}=\pm\pi/2$ and
either $a,b\neq 0$, $c=d=0$ or $c,d\neq 0$, $a=b=0$, therefore
there are totally four degenerate solutions. One can see from Fig.
\ref{VL} that this degeneracy can also be inferred naturally from
the geometry of the vortex configuration.

The most favorable relative phases can also be understood by a
simple Ginzburg-Landau (GL) argument. This GL theory should work
well particularly nearby the phase transition point discussed
below, where the order parameter is small. For those order
parameters that definitely co-exist, we first write down the most
general coupling form between them by momentum conservation, and
then determine the most favorable relative phases by minimizing
energy. For instance, for $p/q=1/3$, $\Delta^0$, $\Delta^1$ and
$\Delta^2$ all co-exist, and then one can write
\begin{eqnarray}
E_{\text{GL}} \propto
\Delta^{0*}\Delta^{0*}\Delta^1\Delta^2+\Delta^{1*}\Delta^{1*}\Delta^0\Delta^2+\Delta^{2*}\Delta^{2*}\Delta^0\Delta^1+\text{c.c},\nonumber
\end{eqnarray}
thus the energy depends on the phases as
$\cos(2\theta_1-\theta_2)+\cos(2\theta_2-\theta_1)+\cos(\theta_1+\theta_2)$,
and one can easily show that the angles listed above are its
minima. For $p/q=1/4$, $\Delta^0$ and $\Delta^2$ definitely
co-exist, thus one shall write down
\begin{eqnarray}
E \propto \Delta^{0*}\Delta^{0*}\Delta^2\Delta^2+\text{c.c},\nonumber
\end{eqnarray} which gives
the energy-phase relative as $\cos 2\theta_1$, whose minima occur
at $\theta_1=\pm \pi/2$.

{\it Vortex Configuration:} To study the configuration of vortices
in the superfluid ground state, we first note that in presence of
magnetic field the Wannier wavefunction at each site should be
chosen as
\begin{equation}
\varphi({\bf r-R_j})=e^{i 2\pi
p(x/\lambda)(y/\lambda-j_y)/q}\varphi^0({\bf r-R_j}),
\end{equation}
where $\varphi^0$ is a Wannier wavefunction in absence of magnetic
field, $\lambda$ is the lattice spacing. The real space profile of the order parameter $\Delta({\bf
r})=\sum_j\Delta_j\varphi({\bf r-R_j})$ is contour-plotted in Fig.
\ref{VL} for two different flux densities. We also verify that the
phase of $\Delta({\bf r})$ winds $2\pi$ around each vortex core.
There are six (four) space group symmetry-related configurations
for Fig. \ref{VL}(a)(b), which corresponds to six (four)
degenerate mean-field solutions. Hence, we
have presented a systematical way to determine the configuration
of vortices in a BCS superfluid from a microscopic theory, which can be verified experimentally with standard imaging technique in cold atom experiments.

{\it Insulator (semi-metal) to superfluid transition:} For
$n_{\text{B}}=p/q$, and for the fermion density of each spin
component $n=\nu/q$, where $\nu$ is an integer from $1,\dots,q-1$,
the system is usually a Hofstadter insulator in absence of
interactions. Except for the case that $q$ is an even integer and
$n=1/2$, the system is a semi-metal since there are Dirac nodes at
the Fermi energy. In both cases, since the Fermi energy is either
in the band gap (Hofstadter insulator), or the density-of-state
linearly vanishes (Hofstadter semi-metal) at the Fermi energy,
there is no Cooper instability for infinitesimally small
attractive interactions. Thus, it requires a critical interaction
strength to turn the system into a paired superfluid through a
second-order phase transition, as shown in Fig. \ref{Phase}(b). In
this calculation, we also fix the fermion density by judging
chemical potential. This transition is driven by the competition
between pairing-energy gain and the single-particle energy cost to
excite particles across the band gap, which was first discussed in
Ref. \cite{Zhai} for a lattice system without magnetic field.
Without the magnetic field, to realize the transition one needs to
tune the interaction close to a Feshbach resonance to achieve
strong pairing strength comparable to the band gap; while in this
case, since the magnetic band gap is controlled by original band
width $t$, the transition can be achieved by varying $U/t$, as
routinely done in cold-atom experiments. This transition is accomplished by a change in compressibility and can be measured directly from {\it in situ} density profile, which has been successfully used in studying boson Hubbard model.

{\it Acknowledgment}: HZ is grateful to Fei Ye for the discussion
of magnetic translation group, and we thank Jason Ho for helpful
correspondence. HZ is supported by the Basic Research Young
Scholars Program of Tsinghua University, NSFC Grant No. 10944002
and 10847002. R.O.U. is supported by T\"{U}B\.{I}TAK. M.O.O is
supported by TUBITAK-KARIYER Grant No. 104T165.


\begin{thebibliography}{99}

\bibitem{review}
For a review, see I. Bloch, J. Dalibard, and W. Zwerger, Rev. Mod.
Phys. {\bf 80}, 885 (2008).

\bibitem{fetter}
The physics of rotating bosons has been recently reviewed in A. L.
Fetter, Rev. Mod. Phys. {\bf 81}, 647 (2009).

\bibitem{Cornell}
V. Schweikhard, {\it et al.}
Phys. Rev. Lett. {\bf 92}, 040404 (2004); S. Tung, V. Schweikhard,
and E. A. Cornell, Phys. Rev. Lett. {\bf 97}, 240402 (2006).

\bibitem{gauge_theory}
G. Juzeli\={u}nas and P. \"Ohberg, Phys. Rev. Lett. {\bf 93},
033602 (2004); G. Juzeli\={u}nas, J. Ruseckas, P. \"Ohberg, and M.
Fleischhauer, Phys. Rev. A {\bf 73}, 025602 (2006); J. Ruseckas,
G. Juzeli\={u}nas, P. \"Ohberg, and M. Fleischhauer, Phys. Rev.
Lett. {\bf 95}, 010404 (2005); E. J. Mueller,  Phys. Rev. A {\bf
70}, 041603 (2004).

\bibitem{NIST}
Y.-J. Lin, {\it et al.}
Nature, {\bf 462}, 628 (2009).

\bibitem{Hofstadter} D. R. Hofstadter, Phys. Rev. B \textbf{14}, 2239 (1976).

\bibitem{BHH}
M. Niemeyer, J. K. Freericks, and H. Monien, Phys. Rev. B {\bf
60}, 2357 (1999); D. Jaksch and P. Zoller, New J Phys. {\bf 5}, 56
(2003); M. \"{O}. Oktel, M. Ni\c{t}\u{a}, and B. Tanatar, Phys.
Rev. B \textbf{75}, 045133 (2007); C. J. Wu, H. D. Chen, J. P. Hu,
and S. C. Zhang, Phys. Rev. A {\bf 69}, 043609 (2004); R. O. Umucal\i lar and M. \"O, Oktel, {\it ibid},
{\bf 76}, 055601 (2007);  D. S. Goldbaum and E. J. Mueller, {\it ibid},
{\bf 79} 021602 (2009); {\it ibid}, {\bf 77} 033629 (2008); K.
Kasamatsu, {\it ibid}, {\bf 79}, 021604 (R) (2009); A. S. S\o rensen, E. Demler, and M. D. Lukin, Phys. Rev. Lett. {\bf 94},
086803 (2005); G. M\"oller
and N. R. Cooper, Phys. Rev. Lett. {\bf 103}, 105303 (2009) and T.
Duric, D. K. K. Lee, Phys. Rev. B {\bf 81}, 014520 (2010).






\bibitem{Zak}
J. Zak, Phys. Rev. {\bf 134}, A1602 (1964), {\it ibid}, {\bf 134},
A1607 (1964), I. Dana, Y. Avron and J Zak, J. Phys. C: Solid State
Phys. {\bf 18}, L679 (1985).

\bibitem{Niu}
For a review, see D Xiao, M. C. Chang and Q. Niu, arXiv: 0907.2021, Rev. Mod. Phys, to be published, Sec VIII.


\bibitem{Zhai}
H. Zhai and T. L. Ho, Phys. Rev. Lett, {\bf 99}, 100402 (2007).

\end{thebibliography}
\end{document}